**Short Paper**

# Dependency on Meta AI Chatbot in Messenger Among STEM and Non-STEM Students in Higher Education

Hilene E. Hernandez
College of Computing Studies, Don Honorio Ventura State University, Philippines
hehernandez@dhvsu.edu.ph
https://orcid.org/0009-0008-9356-8178
(corresponding author)

Rhiziel P. Manalese
College of Arts and Sciences, Don Honorio Ventura State University, Philippines
https://orcid.org/0009-0007-9934-5971
rpmanalese@dhvsu.edu.ph

Roque Francis B. Dianelo
College of Education, Don Honorio Ventura State University, Philippines
https://orcid.org/0009-0000-4673-282X
rfbdianelo@dhvsu.edu.ph

Jaymark A. Yambao
College of Computing Studies, Don Honorio Ventura State University, Philippines
https://orcid.org/0009-0006-5403-1840
jayambao@dhvsu.edu.ph

Almer B. Gamboa
College of Education, Don Honorio Ventura State University, Philippines
https://orcid.org/0009-0008-6281-2764
abgamboa@dhvsu.edu.ph

Lloyd D. Feliciano
College of Hospitality and Tourism Management, Don Honorio Ventura State University, Philippines
https://orcid.org/0009-0009-0191-6356
ldfeliciano@dhvsu.edu.ph

Mike Haizon M. David
College of Hospitality and Tourism Management, Don Honorio Ventura State University, Philippines
https://orcid.org/0000-0001-9285-406X
mhmdavid@dhvsu.edu.ph

Freneil R. Pampo
College of Engineering and Architecture, Don Honorio Ventura State University, Philippines
https://orcid.org/0000-0002-4857-6012
frpampo@dhvsu.edu.ph

John Paul P. Miranda
College of Computing Studies, Don Honorio Ventura State University, Philippines
https://orcid.org/0000-0003-2059-972X
jppmiranda@dhvsu.edu.ph

**Abstract**

*Purpose* – To understand the potential dependency of tertiary students regarding Meta AI in the academic context.

*Method* – This descriptive cross-sectional study surveyed 872 tertiary students from public and private institutions in Luzon, Philippines. Demographic information and perceptions on Meta AI dependency based on existing literature were collected. Descriptive statistics were used to summarize the data and differences between STEM and non-STEM students were analyzed using the Mann-Whitney U test.

*Results* – The results indicate a nuanced perspective on Meta AI chatbot use among students. While there is general disagreement with heavy reliance on the chatbot for academic tasks, psychological support, and social factors, there is moderate agreement on its technological benefits and academic utility. Students value the Meta AI convenience, availability, and problem-solving assistance, but prefer traditional resources and human interaction for academic and social support. Concerns about dependency risks and impacts on critical thinking are acknowledged, particularly among STEM students, who rely more on chatbots for academic purposes. This suggests that while Meta AI is a valuable resource, its role is complementary rather than transformative in educational contexts, with institutional encouragement and individual preferences influencing usage patterns.

*Conclusion* - Students generally hesitate to rely heavily on meta-AI chatbots. This reflects a preference for traditional resources and independent problem-solving. While students acknowledge AI chatbots academic benefits and technological convenience, concerns about overreliance and its impact on critical thinking persist, particularly among STEM students, who appear more inclined to integrate these tools into their studies.

*Recommendations* – Educational institutions should encourage the balanced use of AI chatbots by integrating them as supplementary tools while promoting critical thinking and independent problem-solving skills.

*Practical Implications* – Despite students reporting limited reliance on Meta AI, educators and policymakers must proactively prepare for potential future dependency by developing forward-looking policies and continuously evaluating their impact on education.

## INTRODUCTION

Artificial intelligence (AI) tools are increasingly used in higher education to enhance learning and improve administrative efficiency (George & Wooden, 2023; Ghamrawi et al., 2024; Sajja et al., 2024). Among these, the Meta AI chatbot in Messenger represents a widely accessible tool that provides real-time academic support. This study explores the potential dependency on Meta AI chatbot among STEM and non-STEM students in higher education in Luzon, Philippines. It specifically focuses on its role in a developing country context where accessible and free-to-use technologies are vital. It examines differences in dependency across academic disciplines. This study analyzes key factors such as academic reliance, psychological impacts, social influences, technological accessibility, and behavioral learning patterns. Additionally, the study considers institutional and educational contexts, including how policies and resources affect usage, and evaluates risks of over-reliance, such as diminished critical thinking and learning independence. Furthermore, this study aims to provide insights and add to existing discourse to help educational institutions develop strategies for integrating AI tools that support meaningful learning experiences while addressing ethical and pedagogical challenges.

## LITERATURE REVIEW

AI chatbots offer personalized assistance and have been shown to improve student engagement and academic outcomes. For example, Labadze et al. (2023) found that AI chatbots function as virtual teaching assistants, providing timely support and resources. Similarly, Essel et al. (2022) reported that students in resource-limited settings benefit from chatbot-driven academic support. However, concerns have emerged regarding ethical issues, including plagiarism, misinformation, and algorithmic bias. Kooli (2023) emphasized the need for clear guidelines to address these ethical challenges. For instance, research by Sain and Hebebci (2023) shows that AI chatbots, powered by advanced language models, are reshaping traditional education by delivering human-like responses across academic fields. Bettayeb et al. (2024) demonstrated that such tools enhance learning by offering tailored guidance. However, Darwin et al. (2024) and Fonkam et al. (2024) noted that excessive reliance may reduce students' critical thinking. Zhai et al. (2024) found that dependence on AI dialogue systems can impair cognitive development.

Meta AI's integration into Messenger increases its convenience and accessibility, especially in the Philippines, where Facebook is widely used (Balita, 2024; NapoleonCat, 2024). This availability may lead to habitual or excessive use. Studies have linked frequent AI use to reduced motivation, diminished independent problem-solving, and a decline in academic integrity (Bozkurt et al., 2024; Hasanein & Sobaih, 2023; Keengwe, 2023). These findings highlight the need to explore AI chatbot usage and its broader implications in higher education.

## METHOD

### Research Design

This study employed a descriptive cross-sectional research design, which is suitable for capturing a snapshot of AI dependency among tertiary students at a specific point in time (Hunziker & Blankenagel, 2024). This design allows for the examination of patterns, behaviors, and perceptions related to AI chatbot usage, particularly Meta AI in Messenger, without manipulating any variables. The cross-sectional approach was selected to efficiently gather data from a large population within a limited timeframe and provide insights into current trends and potential areas of concern regarding AI dependency (Zhang et al., 2024). The study was conducted to address the limited research on AI dependency among tertiary students in the Philippines, particularly in Luzon, where the majority of higher education institutions (HEIs) are concentrated. As AI tools like Meta AI become increasingly accessible and integrated into students' academic routines, this research may shed light to explore the potential emerging challenges faced by HEIs.

### Respondents and Sampling

The respondents were tertiary students currently enrolled for the academic year 2024 in both public and private HEIs located on Luzon Island, Philippines. Inclusion criteria required that participants were willing to participate in the study. According to the Philippine Commission on Higher Education's 2023 report, there were 4,443,878 enrolled tertiary students in state, local, private, and other government schools. Using this figure, the required sample size was calculated to be a minimum of 384 respondents. A total of 1,176 responses were collected, but after screening for eligibility, only 872 responses were deemed valid. This reflects a 227.08% response rate based on the required sample size of eligible respondents.

## *Respondents Demographics*

The study involved 872 students. Table 2 shows that the average age was 19.32 years (±6.34), with females comprising 54.7% and males 45.3% of the sample. A majority of respondents were non-STEM students (65.5%), while STEM students accounted for 34.5%. Notably, 75.5% were first-year students, with the remaining distributed across other academic years. Regular students constituted 96.4% of the respondents. Regarding study habits, 39.9% dedicated more than four hours daily to studying, 35.4% studied between two to four hours, and 24.7% less than 2 hours. Internet access was reported as moderate by 56.7% of respondents, excellent by 37.7%, and poor by 5.6%. In terms of Meta AI chatbot usage in Messenger, 62% used it rarely, 26.9% a few times a week, and 11% daily. The majority (77.9%) had been using the chatbot for less than a month. Access was predominantly via smartphones (95.7%). Peer influence on chatbot use was moderate for 62.5% of respondents.

## *Research Instrument*

The research instrument was developed based on existing literature concerning AI chatbots in academic contexts. To ensure validity, the instrument was reviewed by three educational technology experts and pre-tested with 50 tertiary students who were excluded from the actual data collection. The reliability of the instrument was confirmed with a Cronbach's alpha coefficient greater than the acceptable threshold of 0.70. The instrument consisted of two sections: The first section gathered information on age, sex, field of study (STEM or Non-STEM), current enrollment status, year level, type of institution, daily study hours, internet access, usage of Meta AI in Messenger, device used to access it, and peer influence. The second section comprised 27 items distributed across nine domains (see Table 2). A four-point Likert scale was used for responses, with the following interpretation scale: Strongly Disagree: 1.00 – 1.75; Disagree: 1.76 – 2.50; Agree: 2.51 – 3.25; Strongly Agree = 3.26 – 4.00.

Table 1. Demographic profile of the respondents

| Demographic profile | n | % |
|---|---|---|
| Entire group | 872 | |
| Age (Mean ± SD) | 19.32 ± 6.34 | |
| Sex | | |
|   Male | 395 | 45.3% |
|   Female | 477 | 54.7 |
| Field of study | | |
|   STEM | 301 | 34.5 |
|   Non-STEM | 571 | 65.5 |
| Year Level | | |
|   1st year | 658 | 75.5 |
|   2nd year | 6 | 0.7 |
|   3rd year | 187 | 21.4 |
|   4th year | 20 | 2.3 |
|   5th year | 1 | 0.1 |
| Type | | |
|   Regular student | 841 | 96.4 |
|   Irregular student | 31 | 3.6 |
| Study hours per day | | |
|   Less than 2 hours | 215 | 24.7 |
|   2 to 4 hours | 309 | 35.4 |
|   More than 4 hours | 348 | 39.9 |
| Internet access | | |
|   Excellent | 329 | 37.7 |
|   Moderate | 494 | 56.7 |
|   Poor | 49 | 5.6 |
| Frequency of using the Meta AI chatbot in Messenger | | |
|   Daily | 96 | 11 |
|   A few times a week | 235 | 26.9 |
|   Rarely | 541 | 62 |
|   Never | - | - |
| Duration of Usage | | |
|   Less than a month | 679 | 77.9 |
|   1 to 3 months | 135 | 15.5 |
|   More than 3 months | 58 | 6.7 |
| The main device used to access Meta AI | | |
|   Smartphone | 835 | 95.7 |
|   Laptop/ Desktop | 29 | 3.3 |
|   Tablet | 8 | 0.9 |
| Peer Influence on Meta AI Use | | |
|   High | 136 | 15.6 |
|   Moderate | 545 | 62.5 |
|   Low | 191 | 21.9 |

Table 2. The basis for the domains of AI chatbot dependency

| Domains | Definition | Basis |
|---|---|---|
| AI chatbot dependency | It is the reliance on the Meta AI chatbot in Messenger for academic or personal tasks. | (Izadi & Forouzanfar, 2024; Kooli, 2023; Labadze et al., 2023; Niloy et al., 2024; Okonkwo & Ade-Ibijola, 2021) |
| Psychological | It is the emotional and mental influences that drive chatbot use, such as reducing stress or boosting confidence. | (Antony & Ramnath, 2023; Zhang et al., 2024) |
| Academic | It is the use of the chatbot to manage workload, improve performance, and support learning. | (Schei et al., 2024; Zhai et al., 2024) |
| Social | It is the role of peer influence and collaboration in encouraging chatbot usage. | (Schei et al., 2024; Yu & Nazir, 2021) |
| Technological | It is the accessibility, ease of use, and reliability of the chatbot that attract users. | (Labadze et al., 2023; Mohebi, 2024) |
| Behavioral and learning patterns | It is how study habits and learning preferences influence chatbot usage. | (Ma et al., 2024; Schei et al., 2024) |
| Institutional and educational context | It is the effect of institutional policies and resource availability on chatbot reliance. | (Gökçearslan et al., 2024; Zhai et al., 2024) |
| Limitations of human interaction | It is the reduced access to human help that increases dependence on the chatbot. | (Antony & Ramnath, 2023; Zhang et al., 2024) |
| Dependency risks | It is the negative effects of overusing the chatbot, like reduced critical thinking or learning independence. | (Gruenhagen et al., 2024; Kooli, 2023; Samala et al., 2024; Zhai et al., 2024) |

## *Data Collection Procedure*

The data collection process was conducted in November 2024 among tertiary students enrolled in public and private HEIs across Luzon, Philippines. To maximize participation and ensure a diverse sample, several teachers from various HEIs voluntarily assisted in distributing the survey to their students. These educators played a key role in reaching students from different academic disciplines and year levels and contributed to a broader representation of the student population. In addition to institutional distribution, a snowball sampling technique was employed to further extend its reach. Participating students were encouraged to refer the survey to their peers, particularly those known to have experience using Meta AI in Messenger, both within and outside their academic institutions. This approach helped capture a more inclusive range of experiences and perspectives related to AI chatbot usage. The survey was administered online for convenience and wider accessibility due to the high digital engagement among Filipino students. Before participation, informed consent was obtained from all respondents through an embedded consent form within the survey. This form outlined the study's purpose, the voluntary nature of participation, confidentiality assurances, and the right to withdraw at any point based on research ethical principles and the Philippine Data Privacy Act of 2012. Respondents were required to acknowledge and agree to the consent terms before proceeding with the questionnaire to comply with the ethical rules and transparency throughout the data collection process.

## *Data Analysis*

The collected data were analyzed using IBM SPSS Statistics software, version 25.0, to ensure accurate statistical processing. Descriptive statistics, including frequency, percentage, mean, and standard deviation, were utilized to summarize the demographic characteristics of the respondents and their responses to survey items. These measures provided a clear overview of the distribution and central tendencies within the dataset. Additionally, to determine the appropriate statistical tests, a normality test was conducted, which revealed that the data were not normally distributed. As a result, non-parametric statistical tests were employed to analyze differences and relationships within the data. Specifically, the Mann-Whitney U test was used to compare AI dependency levels between STEM and non-STEM students, as this test is suitable for assessing differences between independent groups when the data do not meet the assumptions of normality (MacFarland & Yates, 2016). This analytical approach allowed for robust and reliable interpretation of the data and accurately reflected the students' experiences and perceptions regarding Meta AI usage.

# RESULTS

## *Reliance on Meta AI*

The results indicate that users tend to disagree with being heavily reliant on the Meta AI chatbot in Messenger. They reported disagreeing with frequent use of the chatbot for academic tasks (mean = 2.56, SD = 0.748), moderate unease when unable to access them (mean = 2.13, SD = 0.704), and dependency on chatbots to manage academic challenges (mean = 2.09, SD = 0.766). AI chatbots were seen as a moderately preferred resource for problem-solving (mean = 2.38, SD = 0.758). The overall mean for this domain was 2.29 (Table 3; SD = 0.577).

### *Psychological Benefits of AI Chatbots*

Students in this study tended to agree that Meta AI offers some form of psychological support. Specifically, for them, it can slightly increase confidence (mean = 2.28, SD = 0.724, disagree), help reduce anxiety about complex tasks (mean = 2.40, SD = 0.747, disagree), and moderately provide support during stressful times (mean = 2.46, SD = 0.763, disagree). The overall mean for this domain was 2.38 (Table 3; SD = 0.630).

### *Academic Support from AI Chatbots*

Students generally agree that AI chatbots like Meta AI can provide academic benefits. Students agreed that chatbots can help improve academic performance (mean = 2.55, SD = 0.738), assist in managing workloads (mean = 2.48, SD = 0.717, disagree), and clarify difficult concepts (mean = 2.65, SD = 0.705). The overall mean for this domain was 2.56 (Table 3; SD = 0.629).

### *Social Factors Influencing AI Chatbot Use*

Social factors had a mixed influence on Meta AI use, with students tending to disagree overall. Peers had a moderate influence (mean = 2.41, SD = 0.727, disagree), chatbots were slightly helpful for communication or collaboration (mean = 2.55, SD = 0.723, agree), and users preferred seeking help from peers or instructors rather than AI chatbots (mean = 2.31, SD = 0.780, disagree). The overall mean for this domain was 2.42 (Table 3; SD = 0.597).

### *Technological Aspects of AI Chatbots*

Students generally agree with the positive technological aspects of Meta AI. They found that the chatbot is easy and convenient to use (mean = 2.87, SD = 0.647, agree), moderately reliable and accurate (mean = 2.53, SD = 0.675, agree), and appreciated their constant availability (mean = 2.76, SD = 0.640, agree). The overall mean for this domain was 2.72 (Table 3; SD = 0.537).

### *Behavioral Learning Patterns with AI Chatbots*

Students disagreed that Meta AI aligns closely with their preferred learning style (mean = 2.41, SD = 0.694) and that they avoid solving problems independently when chatbots are available (mean = 2.25, SD = 0.724). They also disagreed that AI chatbots like Meta AI have significantly changed their approach to learning (mean = 2.46, SD = 0.724). The overall mean for this domain was 2.37 (Table 3; SD = 0.608).

## Institutional and Educational Context for AI Chatbot Use

Results also indicate a moderate influence of institutional and educational contexts on AI chatbot use. Respondents disagreed that their institutions actively encourage the use of AI chatbots for academics (mean = 2.32, SD = 0.728), but they slightly agreed that limited access to other resources increases their dependency on chatbots (mean = 2.51, SD = 0.739). The overall mean for this domain was 2.42 (Table 3; SD = 0.637).
.
## Limitations of Human Interaction with AI Chatbots

The results also suggest mixed perceptions regarding the limitations of human interactions compared to AI chatbots like Meta AI. Students agreed that chatbots provide faster responses than human help (mean = 2.77, SD = 0.668), but they disagreed with using chatbots due to the unavailability of tutors or instructors (mean = 2.48, SD = 0.723). Additionally, they disagreed with preferring chatbots over face-to-face interactions for academic issues (mean = 2.17, SD = 0.756). The overall mean for this domain was 2.47 (Table 3; SD = 0.562).

## Dependency Risks Associated with AI Chatbot Use

Results also show a moderate concern about dependency risks linked to Meta AI use. Students agreed that overusing chatbots could reduce their critical thinking skills (mean = 2.69, SD = 0.767) and that reliance on chatbots impacts their ability to learn independently (mean = 2.56, SD = 0.775). However, they disagreed with feeling less capable of solving problems without chatbots (mean = 2.28, SD = 0.710). The overall mean for this domain was 2.51 (Table 3; SD = 0.603).

Table 3. Overall interpretation of individual domains

| Domains | Mean | SD | Interpretation |
|---|---|---|---|
| AI chatbot reliance | 2.29 | 0.58 | Disagree |
| Psychological | 2.38 | 0.63 | Disagree |
| Academic | 2.56 | 0.63 | Agree |
| Social | 2.41 | 0.59 | Disagree |
| Technological | 2.72 | 0.54 | Agree |
| Behavioral and learning patterns | 2.37 | 0.61 | Disagree |
| Institutional and educational context | 2.42 | 0.64 | Disagree |
| Limitations of human interaction | 2.47 | 0.56 | Disagree |
| Dependency risks | 2.51 | 0.60 | Agree |

Table 4 shows the results from a test of differences between STEM and non-STEM students using the Mann-Whitney U statistical test. Only the academic domain ($p = 0.042$) reported a significant difference between STEM and non-STEM students. STEM students (Mean = 459.49) score higher, which suggests they rely more on AI chatbots for academic purposes. Other domains have reported non-significant values. Interestingly, AI chatbot reliance has the closest significance ($p$ = 0.059) but does not reach the threshold ($p <$ 0.05).

Table 4. Mann-Whitney $U$ Test Results for Differences Between STEM and Non-STEM Students

| Domain | Group | N | Mean Rank | Sum of Ranks | Mann-Whitney $U$ | Z-Score | Significant? ($p$) |
|---|---|---|---|---|---|---|---|
| AI chatbot reliance | STEM | 301 | 414.54 | 124,777.00 | 79,326.000 | -1.889 | No ($p$ = 0.059) |
|  | Non-STEM | 571 | 448.08 | 255,851.00 |  |  |  |
| Psychological | STEM | 301 | 434.67 | 130,836.50 | 85,385.500 | -0.159 | No ($p$ = 0.874) |
|  | Non-STEM | 571 | 437.46 | 249,791.50 |  |  |  |
| Academic | STEM | 301 | 459.49 | 138,306.00 | 79,016.000 | -2.033 | Yes ($p$ = 0.042) |
|  | Non-STEM | 571 | 424.38 | 242,322.00 |  |  |  |
| Social | STEM | 301 | 430.37 | 129,540.50 | 84,089.500 | -0.532 | No ($p$ = 0.595) |
|  | Non-STEM | 571 | 439.73 | 251,087.50 |  |  |  |
| Technical | STEM | 301 | 437.70 | 131,747.00 | 85,575.000 | -0.107 | No ($p$ = 0.915) |
|  | Non-STEM | 571 | 435.87 | 248,881.00 |  |  |  |
| Behavioral and learning patterns | STEM | 301 | 427.07 | 128,548.50 | 83,097.500 | -0.821 | No ($p$ = 0.412) |
|  | Non-STEM | 571 | 441.47 | 252,079.50 |  |  |  |
| Institutional and educational context | STEM | 301 | 448.98 | 135,143.50 | 82,178.500 | -1.102 | No ($p$ = 0.270) |
|  | Non-STEM | 571 | 429.92 | 245,484.50 |  |  |  |
| Human interaction limits | STEM | 301 | 454.16 | 136,702.00 | 80,620.000 | -1.536 | No ($p$ = 0.125) |
|  | Non-STEM | 571 | 427.19 | 243,926.00 |  |  |  |
| Dependency risks | STEM | 301 | 454.68 | 136,859.00 | 80,463.000 | -1.584 | No ($p$ = 0.113) |
|  | Non-STEM | 571 | 426.92 | 243,769.00 |  |  |  |

| Domain | Group | N | Mean Rank | Sum of Ranks | Mann-Whitney $U$ | Z-Score | Significant? ($p$) |
|---|---|---|---|---|---|---|---|
| | STEM | | | | | | |

# DISCUSSION

Students did not show strong reliance on Meta AI for academic tasks. They used the chatbot when necessary but preferred traditional resources and independent study. This reflects a cautious approach to AI integration in learning. The low mean scores across reliance-related items confirm that students have not formed habits of frequent or dependent use. These results support the argument by Fabio et al. (2024) that maintaining learner autonomy remains important when introducing AI tools in education.

Students also showed limited agreement with statements about psychological support from Meta AI. They did not view the chatbot as a source of emotional comfort, motivation, or anxiety relief. Although Dekker et al. (2020) found that chatbots can reduce stress, the current findings suggest that students still rely more on human support systems for emotional needs. They see AI tools as functional rather than affective. The results show that students understand the boundaries of AI support in terms of emotional and psychological well-being.

Meta AI provided academic support, especially in clarifying difficult concepts and improving performance. Students recognized its usefulness in helping with tasks, though they did not strongly agree that it helped manage academic workloads. Social influence had little impact. Students preferred direct help from peers or instructors instead of relying on the chatbot due to peer encouragement. This supports the findings of Ayanwale and Molefi (2024), who argued that personal attitudes are stronger predictors of technology use than social influence. The results show that academic benefits matter more than social factors when students decide to use AI tools.

STEM students reported higher academic reliance on Meta AI than non-STEM students. The results from the statistical test confirmed this difference. This supports earlier findings by Ma et al. (2024) and Xu and Ouyang (2022), who found that STEM students often use AI to help with technical tasks. Other domains did not show significant differences between the two groups. This suggests that personal preferences, digital literacy, and familiarity with other tools like ChatGPT may affect usage more than discipline alone. Institutions must consider how to provide balanced access and support for AI tools across all programs.

# CONCLUSION AND RECOMMENDATIONS

This study suggests that tertiary students have varied attitudes toward using Meta AI chatbots in Messenger. They generally have low reliance across most domains presented in this study. Students in this study moderately acknowledged the technological advantages, such as ease of use and reliability, yet showed limited alignment of AI chatbots with their learning styles. This reflects a preference for traditional resources and independent learning. However, significant differences were observed between STEM and non-STEM students in the academic domain, where STEM students demonstrated higher reliance on AI chatbots. This aligns with existing literature highlighting the value of AI chatbots for tasks like coding, writing, and immediate feedback in STEM disciplines. The near-significant difference in overall reliance further suggests that STEM students may find these tools particularly useful. This is possibly due to the technical nature of their studies and the alignment of AI capabilities with their academic needs.

In other domains, such as psychological support, technological usability, and social influence, no significant differences were observed, indicating that chatbot usage is influenced by broader factors such as personal preferences, technological proficiency, and the availability of more widely recognized tools like ChatGPT. These findings highlight the need for a balanced integration of AI chatbots to ensure their potential is maximized in enhancing academic outcomes, particularly in STEM. This, in addition to mitigating risks of overreliance and fosters critical thinking and self-directed learning. Institutional efforts should focus on promoting equitable access and aligning AI tools with the diverse needs of students across disciplines.

## PRACTICAL IMPLICATIONS

This study has several practical implications for higher education in the Philippines, where platforms like Facebook and Messenger, including Meta AI, have a dominant presence due to their widespread accessibility and integration into daily communication. The Philippine education system faces challenges such as resource limitations, large class sizes, uneven access to quality educational materials, and disparities between urban and rural areas. In this context, while students viewed limited reliance on Meta AI, its ease of use, constant availability, and moderate academic and psychological benefits suggest it can serve as a valuable supplementary tool, especially where educational resources are scarce. Educational institutions in the Philippines should promote a balanced integration of AI, particularly on responsible AI usage, through targeted AI literacy programs. Given the high mobile and social media penetration rates, particularly among Filipino youth, schools can leverage these platforms to enhance learning while implementing structured policies and ethical guidelines to mitigate risks such as diminished critical thinking and over-reliance. Carefully balancing the use of AI chatbots involves encouraging students to view these tools as supportive aids rather than primary sources of knowledge and actively promoting reflective learning practices where AI-generated content is critically evaluated, and incorporating activities that require independent problem-solving and analytical thinking.

Moreover, despite students viewing themselves as having limited reliance on Meta AI as reported in this study, educators, policymakers, and institutions must proactively prepare for the potential increase in dependency as AI technologies continue to evolve and become more integrated into educational ecosystems. The rapid advancements in AI capabilities, coupled with their growing accessibility, may lead to increased student reliance in the future, which can potentially impact academic integrity, critical thinking, and learning autonomy. Therefore, developing forward-looking policies, integrating AI literacy into curricula, and providing continuous professional development for teachers will help ensure that both students and educators are equipped to manage AI tools effectively. Fostering human-AI collaboration, where AI supports but does not replace human instruction, can optimize learning outcomes in line with these educational goals. Investments in digital infrastructure, equitable resource accessibility, and continuous evaluation of AI's impact on learning behaviors are crucial to ensure that AI tools like Meta AI enhance educational experiences without undermining academic integrity and cognitive development, particularly in resource-constrained or rural areas where traditional academic support may be less accessible.

## DECLARATIONS

### *Conflict of Interest*

The authors declare that there is no conflict of interest related to this study.

### *Informed Consent*

All respondents in this study provided their informed consent by agreeing to the statements about the study as outlined in the data collection questionnaire.

### *Ethics Approval*

Formal ethics approval was not sought for this study as the survey was voluntary. Nevertheless, ethical principles based on the Philippine data privacy law and the Philippine Health Research Ethics Board were rigorously adhered to, including informed consent, data privacy, and respectful conduct.

## Authors Biography

**Hilene E. Hernandez, Rhiziel P. Manalese, Lloyd D. Feliciano**, **Mike Haizon M. David**, and **Freneil R. Pampo** are faculty members at Don Honorio Ventura State University (DHVSU) Main Campus in Bacolor, Pampanga, Philippines.

**Roque Francis B. Dianelo, Jaymark A. Yambao, Almer B. Gamboa,** and **John Paul P. Miranda** are faculty members at DHVSU Mexico Campus in Mexico, Pampanga, Philippines.